\documentstyle[12pt,fleqn]{article}

\font\twlgot =eufm10 scaled \magstep1
\font\egtgot =eufm8
\font\sevgot =eufm7
\font\twlmsb =msbm10 scaled \magstep1
\font\egtmsb =msbm8
\font\sevmsb =msbm7

\newfam\gotfam

\textfont\gotfam\twlgot
\scriptfont\gotfam\egtgot
\scriptscriptfont\gotfam\sevgot

\newfam\msbfam
\textfont\msbfam\twlmsb
\scriptfont\msbfam\egtmsb
\scriptscriptfont\msbfam\sevmsb
\def\Bbb{\protect\pBbb}
\def\pBbb{\relax\ifmmode\expandafter\Bb\else\typeout{You cann't use
Bbb in text mode}\fi}
\def\Bb #1{{\fam\msbfam\relax#1}}

\def\thebibliography#1{\section*{References}\list
   {[\arabic{enumi}]}{\settowidth\labelwidth{#1}\leftmargin\labelwidth
     \advance\leftmargin\labelsep
     \usecounter{enumi}}
     \def\newblock{\hskip .11em plus .33em minus .07em}
     \sloppy\clubpenalty4000\widowpenalty4000
     \sfcode`\.=1000\relax}

\def\op#1{\mathop{\fam0 #1}\limits}

\newcommand{\beq}{\begin{equation}}
\newcommand{\eeq}{\end{equation}}
\newcommand{\ben}{\begin{eqnarray}}
\newcommand{\een}{\end{eqnarray}}
\newcommand{\be}{\begin{eqnarray*}}
\newcommand{\ee}{\end{eqnarray*}}
\newcommand{\bea}{\begin{eqalph}}
\newcommand{\eea}{\end{eqalph}}

\newcommand{\la}{\lambda}

\newcommand{\g}{\gamma}

\newcommand{\ol}{\overline}
\newcommand{\dr}{\partial}

\newcounter{eqalph}
\newcounter{equationa}
\newcounter{theorem}
\newcounter{remark}
\newcounter{proposition}
\newcounter{lemma}
\newcounter{corollary}
\newcounter{definition}

\newenvironment{eqalph}{\stepcounter{equation}
\setcounter{equationa}{\value{equation}}
\setcounter{equation}{0}

\begin{eqnarray}}{\end{eqnarray}\setcounter{equation}{\value{equationa}}}

\def\theremark{\arabic{remark}}

\def\thedefinition{\arabic{definition}}

\newcommand{\mar}[1]{}

\begin{document}
\hbox{}

{\parindent=0pt

{\large\bf The Lyapunov stability of first order dynamic equations
with respect to time-dependent Riemannian metrics. An example.}
\medskip

{\sc Gennadi 
Sardanashvily\footnote{E-mail address: sard@grav.phys.msu.su}}
\bigskip

\begin{small}
Department of Theoretical Physics, Physics Faculty, Moscow State
University, 117234 Moscow, Russia
\bigskip

{\bf Abstract.}
A simple example that I have been requested illustrates the statement
in the E-print nlin.CD/0201060 that solutions of a 
smooth first order dynamic equation can be made
Lyapunov stable at will by the choice of an
appropriate time-dependent Riemannian metric.
\end{small}
}
\bigskip
\bigskip

Let $\Bbb R$ be the time axis provided with the Cartesian coordinate $t$.
In geometric terms, a (smooth) first order dynamic
equation in non-autonomous 
mechanics is defined as
a vector field $\g$ on a smooth fibre bundle
$Y\to\Bbb R$ 
which obeys the condition
$\g\rfloor dt=1$. With respect to bundle coordinates $(t,y^k)$ on $Y$,
this vector field reads
\mar{ch3}\beq
\g=\dr_t + \g^k\dr_k. \label{ch3}
\eeq
The associated first order dynamic
equation takes the form
\be
\dot y^k=\g^k(t,y^j)\dr_k,
\ee
where $(t,y^k,\dot t,\dot y^k)$ are holonomic coordinates on the
tangent bundle $TY$ of $Y$.
Its solutions are trajectories of the vector field
$\g$ (\ref{ch3}).

Let a fibre bundle $Y\to\Bbb R$ be provided with a Riemannian fibre metric
$g$, defined as a section of the symmetrized tensor product
$\op\vee^2 V^*Y\to Y$ 
of the
vertical cotangent bundle $V^*Y$ of $Y\to\Bbb R$. With respect to the
holonomic coordinates $(t,y^k,\ol y_k)$ on $V^*Y$,
it takes the coordinate form
\be
g=\frac12 g_{ij}(t,y^k)\ol dy^i\vee \ol dy^j, 
\ee
where $\{\ol dy^i\}$ are the holonomic fibre bases for $V^*Y$.

With a Riemannian fibre metric $g$, 
the instantwise distance $\rho_t(s,s')$
between two solutions $s$ and $s'$
of a dynamic equation $\g$ on $Y$ at an instant $t$
is defined as the distance between the points $s(t)$ and $s'(t)$
with respect to the Riemannian metric $g(t)$ on the fibre $Y_t$ over
$t\in\Bbb R$.
Then the (upper) Lyapunov exponent of a
solution $s'$ with respect to
a solution $s$ is the limit
\mar{ch51'}\beq
K(s,s')=\op{\ol{\lim}}_{t\to\infty} \frac{1}{t}
\ln(\rho_t(s,s')). \label{ch51'}
\eeq
If the upper limit
\be
\op{\ol{\lim}}_{\rho_{t=0}(s,s')\to 0}K(s,s')=\la
\ee
is negative, the solution $s$ is said to be exponentially Lyapunov
stable. If there exists at least one positive Lyapunov exponent, one speaks
about chaos in a dynamical system. 

Proposition 6 in E-print nlin.CD/0201060 states the following.

\begin{itemize}
\item Let $\la$ be a real number. Given a dynamic equation 
defined by a complete vector field $\g$ (\ref{ch3}),
there exists a Riemannian fibre metric on $Y$ such that
the Lyapunov spectrum of any solution of $\g$ is $\la$.
\end{itemize}
The following example aims to illustrate this fact.

Let us consider one-dimensional motion on the axis $\Bbb R$ defined by
the first order dynamic equation 
\mar{y1}\beq
\dot y=y \label{y1}
\eeq
on the fibre bundle 
$Y=\Bbb R\times \Bbb R\to \Bbb R$ coordinated by $(t,y)$. 
Solutions of the equation (\ref{y1}) read
\mar{y2}\beq
s(t)=c\exp(t), \qquad c={\rm const}. \label{y2}
\eeq
Let $e_{yy}=1$ be the standard Euclidean metric on $\Bbb R$. With respect to
this metric, the instantwise distance between two arbitrary solutions 
\mar{y4}\beq
s(t)=c\exp(t), \qquad s'(t)=c'\exp(t) \label{y4}
\eeq
of the equation (\ref{y1}) is
\be
\rho_t(s,s')_e=|c-c'|\exp(t).
\ee
Hence, the Lyapunov exponent $K(s,s')$ (\ref{ch51'}) equals 1,
and so is the Lyapunov spectrum of any solution (\ref{y2}) of the first
order dynamic equation (\ref{y1}). 

Let now $\la$ be an arbitrary real number. In accordance with
Proposition 1 in E-print nlin.CD/0201060, there exists a coordinate
\be
y'=y\exp(-t)
\ee
on $\Bbb R$ such that, written relative to this coordinate, the solutions 
(\ref{y2}) of the equation (\ref{y1}) read $s(t)=$const.
Let us choose the Riemannian fibre metric on $Y\to\Bbb R$ which takes the form
\be
g_{y'y'}=\exp(2\la t) 
\ee
with respect to the coordinate $y'$. Then relative to
the coordinate $y$, it reads
\mar{y3}\beq
g_{yy}=\frac{\dr y'}{\dr y}\frac{\dr y'}{\dr y}g_{y'y'}=\exp(2(\la-1)t).
\label{y3}
\eeq
The instantwise distance between the solutions $s$ and $s'$ (\ref{y4})
with respect to the metric $g$ (\ref{y3}) is 
\be
\rho_t(s,s')=[g_{yy}(s(t)-s'(t))^2]^{1/2}=|c-c'|\exp(\la t).
\ee
One at once obtains that the Lyapunov spectrum of any solution of the
differential equation (\ref{y1}) with respect to the metric (\ref{y3})
is $\la$.

\end{document}